\documentclass{article}
\usepackage{spconf,amsmath,graphicx}
\usepackage{booktabs}
\usepackage{multirow}
\usepackage{amssymb}
\usepackage{amssymb}% http://ctan.org/pkg/amssymb
\usepackage{pifont}
\newcommand{\cmark}{\ding{51}}%
\newcommand{\xmark}{\ding{55}}%
\usepackage{color}

\usepackage{multirow}
\usepackage{subfigure}

{

\title{Audio-visual Recognition of Overlapped speech for the LRS2 dataset}
\address{$^1$The Chinese University of Hong Kong \\ $^2$Tencent AI Lab, Bellevue, USA,~~~ $^3$Tencent AI Lab, Shenzhen, China\\ $^4$Center of Robust Speech Systems (CRSS), University of Texas at Dallas\\
}
\name{Jianwei Yu$^{1,2,*}$\thanks{$^*$ This work was done while the author was an intern at Tencent AI.}, Shi-Xiong Zhang$^2$, Jian Wu$^3$,  Shahram Ghorbani$^4$, Bo Wu$^3$, Shiyin Kang$^3$, \\Shansong Liu$^1$, Xunying Liu$^1$, Helen Meng$^1$, Dong Yu$^{2,3}$}

\begin{document}
\ninept
\maketitle
\begin{abstract}
Automatic recognition of overlapped speech remains a highly challenging task to date. Motivated by the bimodal nature of human speech perception, this paper investigates the use of audio-visual technologies for overlapped speech recognition. Three issues associated with the construction of audio-visual speech recognition (AVSR) systems are addressed. First,  the basic architecture designs i.e. end-to-end and hybrid of AVSR systems are investigated. Second, purposefully designed modality fusion gates are used to robustly integrate the audio and visual features. Third, in contrast to a traditional pipelined architecture containing explicit speech separation and recognition components, a streamlined and integrated AVSR system optimized consistently using the lattice-free MMI (LF-MMI) discriminative criterion is also proposed. 
The proposed LF-MMI time-delay neural network (TDNN) system establishes the state-of-the-art for the LRS2 dataset. 
Experiments on overlapped speech simulated from the LRS2 dataset suggest the proposed AVSR system outperformed the audio only baseline LF-MMI DNN system by up to 29.98\% absolute in word error rate (WER) reduction, and produced recognition performance comparable to a more complex pipelined system. Consistent performance improvements of 4.89\% absolute in WER reduction over the baseline AVSR system using feature fusion are also obtained.

\end{abstract}
\section{Introduction}
\label{sec:intro}

% ==========================================================
% target: Overlapped Speech Recognition
%       1. Why Muti-Modality
%       2. Why integrated
%       3. Fusion methods
% ==========================================================

%%%%%%%%%%%%%%%%%%%%%%%%%%%%%%%%%%%%%%%%%%%%%%%%%%%%%%%%%%%%%%%%%%%%
%
%
%
%
%%%%%%%%%%%%%%%%%%%%%%%%%%%%%%%%%%%%%%%%%%%%%%%%%%%%%%%%%%%%%%%%%%%%%

% 感觉读起来更像一篇journal的introduction
% 
% 相对自己之前写的 introduction
% 文章的对于相关的前沿工作挖掘的更深更全面。 (Yu Jianwei 对分离技术的描述极度肤浅，基本只讲了框架)
% contribution part 缺少了与前人工作的对比。 (Yu Jianwei 未能指出第一次在Overlap AVSR上使用LFMMI这个点)
% 逻辑链：
% andrew: Overlapped speech 很难 -> 现有的分离方法 -> 多模态方法的对于 Overlap speech 研究的动机 -> 该方法的需要解决的问题 -> 文章中解决这些问题的方法 -> 文章的主要贡献
% Jianwei: Overlapped speech 很难 -> 现有的分离方法 (只讲了存在的问题)-> 多模态方法的对于 Overlap speech 研究的动机 -> (缺少了对于问题的描述) 文章中解决这些问题的方法 -> 文章的主要贡献

% Overlapped recognition 为什么很难
Automatic speech recognition (ASR) of overlapped speech is a highly challenging task to date. The presence of interfering speakers introduces a large mismatch between clean and overlapped speech and significant performance degradation. To this end, previous research efforts were heavily focused on speech separation techniques that can convert mixed speech into speaker dependent signals. 

% 关于overlapped speech 的语音分离方法：
% 1. 深度学习之前的 “声场景分析法” 
% 2. 深度学习的 说话人顺序检测 和 双说话人解码方法
% 3. 深度聚类方法
% 4. PIT
% 5. 多通道的波束形成算法
Prior to the deep learning era, computational auditory scene analysis (CASA) \cite{rouat2008computational} approaches containing perceptual cue \cite{wertheimer1938laws} based time-frequency mixed speech decomposition and grouping stages were often used. Amid the rapid progress brought by deep learning technologies to speech recognition, they have drawn increasing research interests for overlapped speech separation and recognition. Deep neural network (DNN) based speaker turn detection and weighted finite-state transducer (WFST) two-talker decoding approaches were proposed for single channel multi-talker speech recognition in \cite{weng2015deep}. Deep clustering based separation techniques that use spectrogram embeddings were proposed in \cite{hershey2016deep,chen2017deep,isik2016single}. Permutation invariant training (PIT) \cite{yu2017permutation} was also developed as a general solution to map single channel, monaural mixed speech inputs to those of individual speakers. When multi-channel microphone arrays are employed, acoustic beaming algorithms \cite{pados2001iterative} or neural network based beamforming architectures \cite{yoshioka2018recognizing} can be used to enhance the desired speaker’s signal, while attenuating the interference from other speakers.

% 使用AVSR方法的动机：人的多感知特性，视觉信息不受音频噪声干扰， 没多少人做
Motivated by the bimodal nature of human speech perception \cite{wertheimer1938laws,mcgurk1976hearing}, and the invariance of visual information to acoustic signal corruption, audio-visual speech recognition (AVSR) technologies \cite{afouras2018deep,zhang2019robust,mroueh2015deep,noda2015audio} can also be used for overlapped speech separation \cite{khan2013speaker, ephrat2018looking,afouras2018conversation,lu2018listen,lu2019audio,khan2018using, wu2019time,khan2013speaker2} and the back-end recognition component. However, the use of visual modality in the recognition stage of system development for overlapped speech remains limited to date. To the best of our knowledge, the only previous work in this direction was reported in \cite{chao2019speaker}.

% AVSR 方法的问题： 1. 模态集成的问题 2. 级联模型训练准侧不一致
Three issues need to be addressed when developing such systems.
First, the fundamental architecture design of AVSR system i.e. end-to-end and hybrid needs to be investigated. This issue has been studied recently in ASR \cite{luscher2019rwth}, while it is still remain unclear for AVSR to date.
Second, in order to robustly integrate the audio and visual modalities, a careful design of modelling components for modality fusion is required. Traditionally a simple audio-visual feature concatenation based fusion scheme can be used \cite{chao2019speaker}. However, it provides limited flexibility in fusion when the visual inputs become less reliable and poorer in quality \cite{potamianos2003recent,DAVSR,tao2018gating}. Third, state-of-the-art systems developed for overlapped speech recognition are often based on a pipelined architecture containing explicit speech separation and recognition components \cite{du2014robust,afouras2019my,afouras2018conversation}. The front-end separation components are often optimized using error costs that are different from those used in the back-end recognition components. Moreover, they often require parallel training data containing clean speech as the learning targets, which are impractical to obtain for unseen and potentially mismatched domains or tasks. This can further limit such systems’ wider application.

% acoustic hidden feature gate driven by visual modality only
%acoustic hidden feature gate driven by audio-visual modalities
% 针对这些问题提出了哪些方法：1. 门方法来集成模态； 2. 统一系统训练准则
In order to address these issues, a comparison between hybrid system and end-to-end systems is first performed to find a strong basic architecture of AVSR systems. Two gated neural architectures are used to facilitate a dynamic fusion between the audio and visual modalities. A streamlined and integrated AVSR system architecture containing implicit speech enhancement and recognition components optimized consistently using the lattice-free MMI (LF-MMI) discriminative criterion is also proposed. Consistent with ASR \cite{luscher2019rwth}, the hybrid system show better performance over published end-to-end\cite{afouras2018deep,petridis2018audio} systems of AVSR on LRS2\cite{chung2017lip} dataset.   
Experiments on overlapped speech simulated from the LRS2 dataset suggest the proposed gated AVSR systems outperforms the audio only baseline LF-MMI time-delay neural network (TDNN) system by up to 29.9\% absolute (74\% relative) in WER reduction, and produced recognition performance comparable to a baseline pipelined AVSR system with a more complex speech separation component. Performance improvements of 4.89\% absolute (32\% relative) in WER reduction over the baseline AVSR system using feature concatenation based modality fusion are also obtained.

% 文章的主要贡献点：1. 门方法来集成模态； 2. 统一系统训练准则
% 可能会被 argue 的地方：1. gate 在音视频语音分离中极度常用
% 对比了前人的工作
The main contributions of this paper includes: 1) this paper is one of the beginning works to compare the performance of hybrid and end-to-end architectures of AVSR, our LF-MMI TDNN system shows state-of-the-art performance on LRS2 dataset; 2) to the best of our knowledge, this paper is the first work using a gated neural network architecture to robustly integrate audio and visual modalities for overlapped speech recognition. In contrast, the only known previous AVSR research for overlapped speech used feature concatenation based fusion \cite{chao2019speaker}; 3) this paper is also the first attempt to use the LF-MMI discriminative criterion to train an integrated AVSR system for overlapped speech. In previously research reported in \cite{chen2018progressive}, the LF-MMI criterion was used in a jointly trained pipelined system with audio inputs only, a more complicated system architecture, training procedure and explicit requirement of parallel training data for constructing the separation component. 

The rest of this paper is organized as follows. Section 2 reviews pipelined audio-visual separation and recognition baseline systems. Section 3 proposes modality fusion gates based AVSR system architectures.  Experiments and results are presented in section 4. Section 5 draws the conclusion and discusses future work.

\begin{figure}[htb]
\begin{minipage}[b]{1\linewidth}
  \centering
  \centerline{\includegraphics[width=9cm]{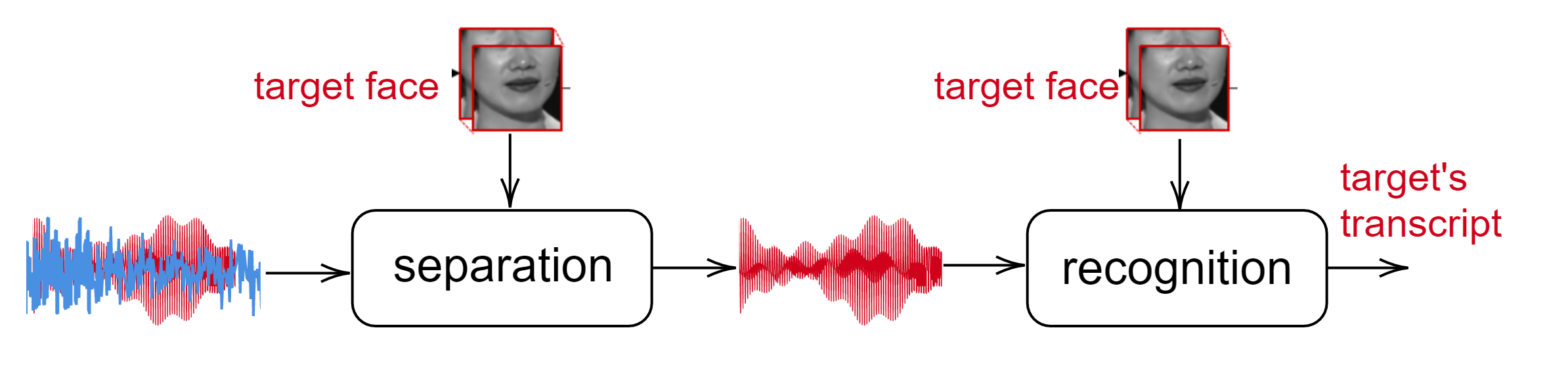}}
  \centerline{(a) Pipelined AVSR system}\medskip
%  \vspace{2.0cm}
\end{minipage}
\begin{minipage}[b]{1\linewidth}
  \centering
  \centerline{\includegraphics[width=9cm]{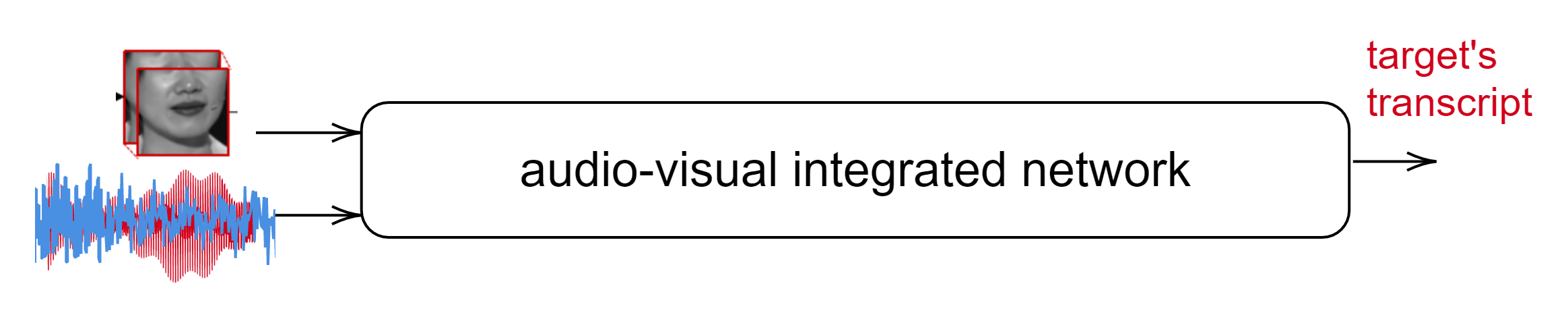}}
  \centerline{(b) Integrated AVSR system}\medskip
%  \vspace{2.0cm}
\end{minipage}
\caption{Illustration of pipelined and integrated  AVSR systems.}
\label{figures}
\end{figure}

% \begin{figure}[t]
% \begin{minipage}[b]{1\linewidth}
%   \centering
%   \centerline{\includegraphics[width=9cm]{Freq-Separation.pdf}}
% %  \vspace{2.0cm}
% \end{minipage}
% \caption{Illustration of audio-visual separation system. The LipNet structure is shown in Figure 2 (a). The structure of the Separation can be found in \cite{wu2019time}}
% \label{figures}
% \end{figure}

\section{Pipelined AVSR for overlapped speech}
This section introduces the architecture of the pipelined AVSR system.
The audio-visual separation component and recognition component are introduced in section 2.1 and Section 2.2 respectively.

\subsection{Audio-visual speech separation}
\label{sssec:subhead}
Recent research on leveraging visual modality has led to impressive results in speech separation.
In these studies, various representations of the visual information such as 
lip appearance \cite{afouras2018conversation,ephrat2018looking} and optical flow\cite{lu2018listen, lu2019audio} are used to estimate the time-frequency (TF) mask. In this paper, the audio-visual speech separation\footnote[1]{Model details, Si-SNR evaluation and audio samples can be found in: \\ https://yjw123456.github.io/Audio-visual-Overlapped-speech-recognition} component used in the pipelined system is based on our previous work in  \cite{wu2019time}. Given an overlapped audio signal and the target speakers' mouth region of interest (ROI). The system uses the visual information to bias the separation network to directly estimate the TF mask of the target speaker. Then the spectrogram of the separated audio is obtained by multiplying the TF mask with the overlapped spectrogram.

\subsection{Audio-visual speech recognition}
\label{sssec:subhead}
For visual modality is still complementary to the separated (enhanced) audio, AVSR system is used as the recognition component in our pipelined system. In the recent studies, the AVSR systems are largely based on end-to-end architectures, such as attention-based encoder-decoder \cite{chung2017lip,afouras2018deep}, Connectionist-Temporal-Classfication (CTC) \cite{afouras2018deep} and hybrid CTC/attention \cite{petridis2018audio}.
% In the recent studies, the AVSR systems are largely based on end-to-end architectures, such as attention-based encoder-decoder systems concatenating the context vectors of individual modality \cite{chung2017lip} 
% You should find a better way to introduce the previous end to end work.
% In the recent studies, the AVSR systems are largely based on end-to-end architectures.
% In \cite{chung2017lip}, the watch, listen, attend and spell (WLAS) model concatenates two context vectors obtained by attending over individual modality to predict output units is proposed. CTC transformer (TM-CTC) concatenates visual and audio features (early modality fusion) and sequence-to-sequence transformer (TM-seq2seq) concatenates visual and audio context vectors are proposed in \cite{afouras2018deep}. Hybrid CTC/attention architecture using early modality fusion is also investigated in \cite{petridis2018audio}.
Motivated by the impressive results of hybrid system in ASR \cite{luscher2019rwth}, in this paper, we investigate the hybrid TDNN AVSR system. 
The structure of the hybrid model used in the pipelined system is shown in Figure 2 (a). The mouth ROI of the target speaker is fed into the LipNet to generated the visual features. The RecogNet is a TDNN network with factored time-delay neural network (TDNN-F) \cite{tdnnf} components,  which has been shown to be effective in modeling long range temporal dependencies \cite{tdnnf}. In our experiments, the hybrid TDNN AVSR system trained with LF-MMI criterion demonstrates the state-of-the-art performance on the LRS2 dataset. Due to the achieved good results, we use the hybrid structure in the rest of this paper.

\begin{figure*}[htb]
\begin{minipage}[b]{0.33\linewidth}
  \centering
  \centerline{\includegraphics[width=6.5cm]{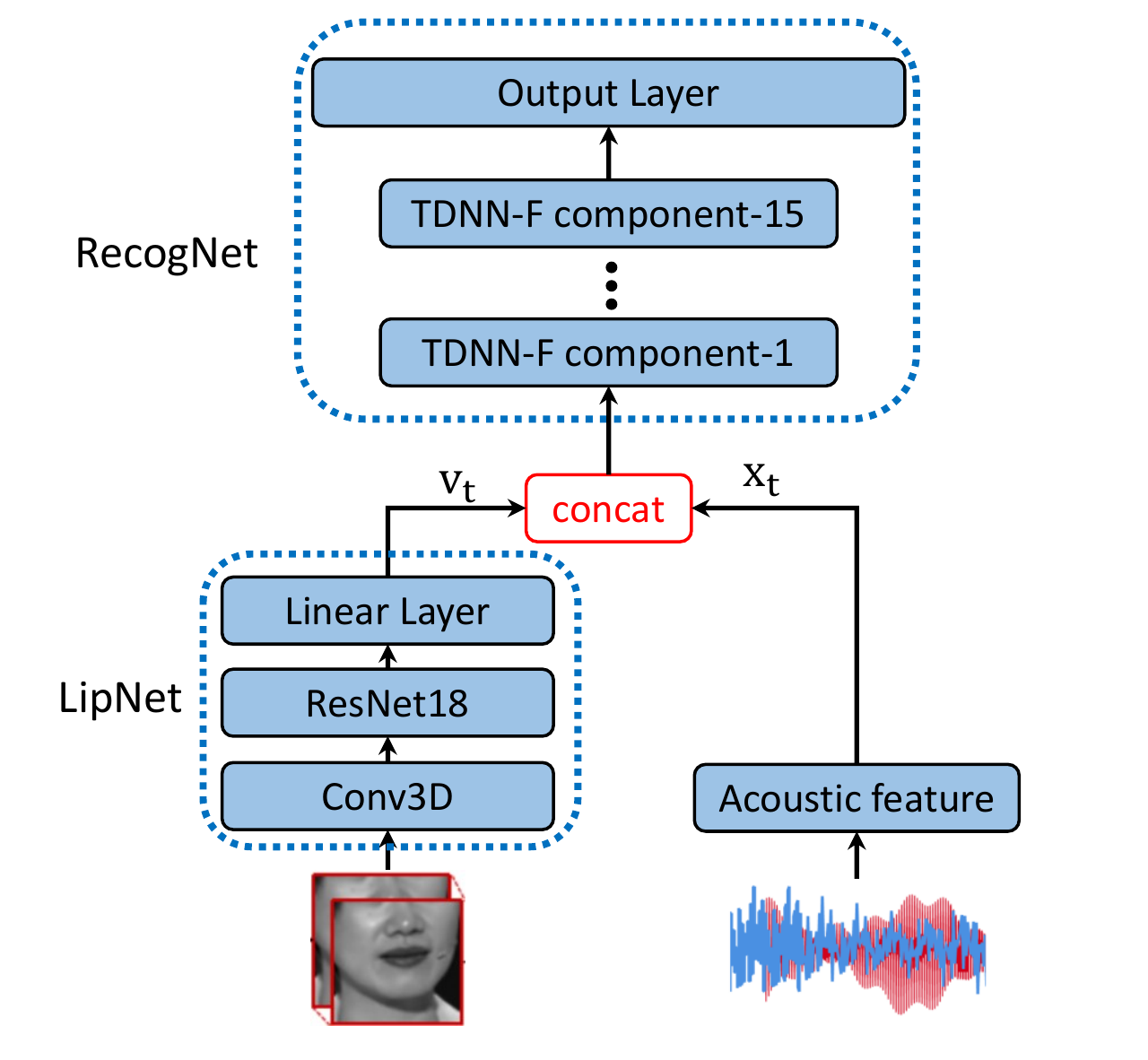}}
%  \vspace{2.0cm}
  \centerline{(a) Feature concatenation}\medskip
\end{minipage}
\begin{minipage}[b]{0.33\linewidth}
  \centering
  \centerline{\includegraphics[width=6.5cm]{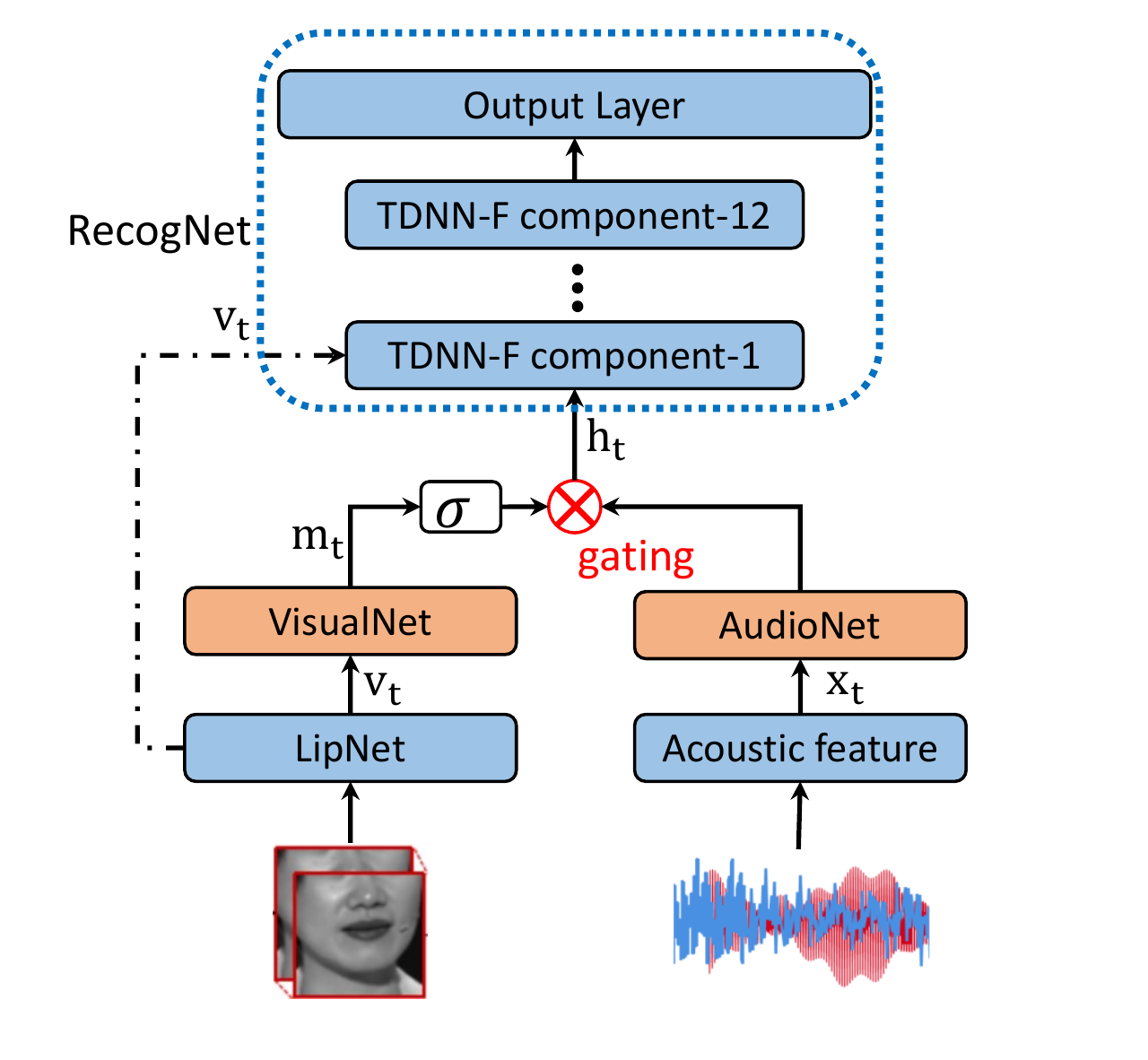}}
%  \vspace{1.5cm}
  \centerline{(b) Visual modality driven gate}\medskip
\end{minipage}
\begin{minipage}[b]{0.33\linewidth}
  \centering
  \centerline{\includegraphics[width=6.5cm]{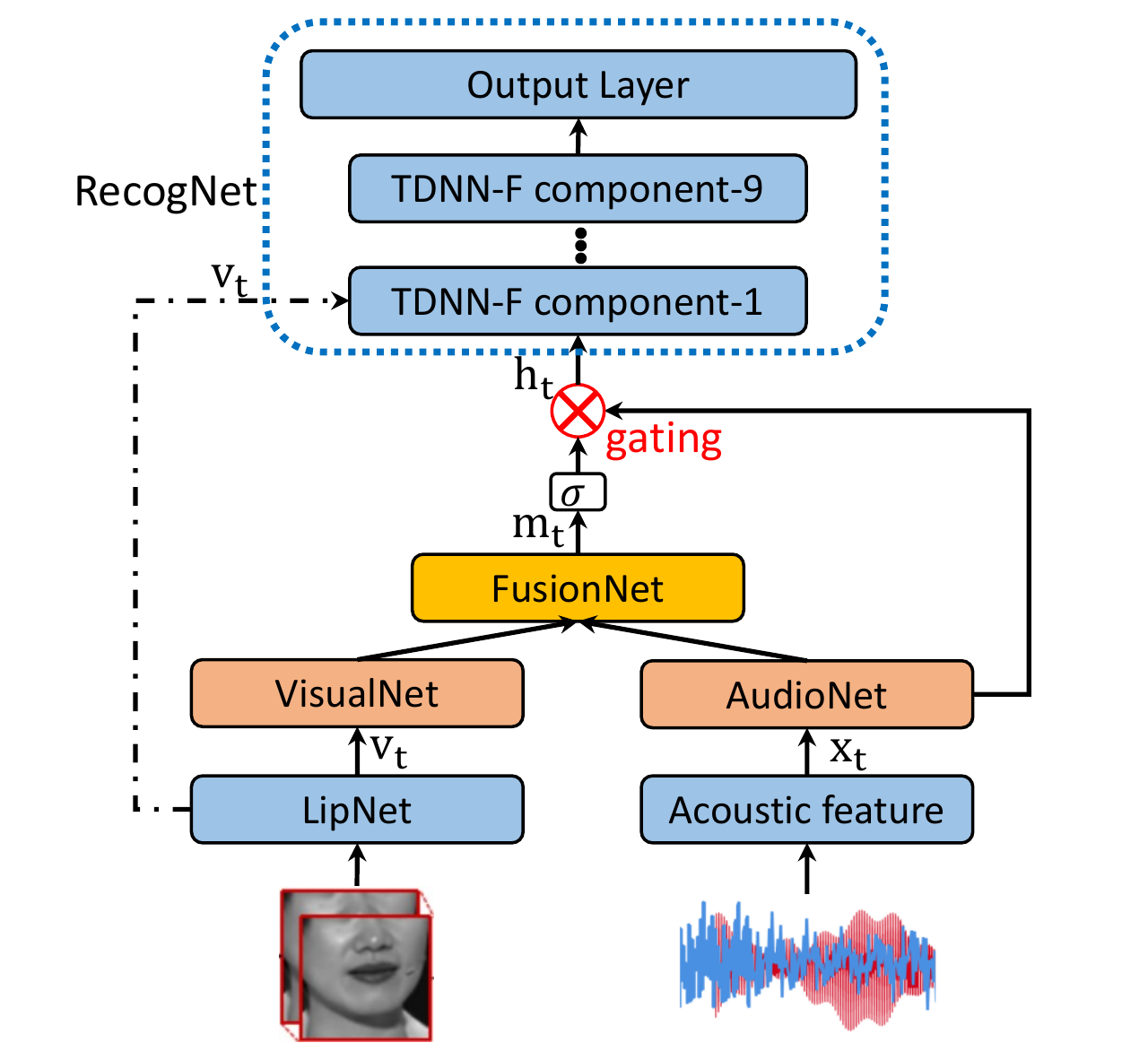}}
%  \vspace{1.5cm}
  \centerline{(c) Audio-visual modality driven gate}\medskip
\end{minipage}
\caption{Illustration of audio-visual fusion methods for AVSR systems: (a) feature concatenation based fusion: acoustic and video features are concatenated before fed into the RecogNet; (b) visual modality driven gated fusion; (c) audio-visual modality driven gated fusion. "{\color{red}$\otimes$}" denotes Hadamard product. The dashed arrow denotes concatenating the gated hidden outputs with the visual features.}
\label{figures}
\end{figure*}

\section{Integrated AVSR for overlapped speech}
\label{sec:format}
The modality fusion methods and the integrated system architecture are introduced in this section.

\subsection{Audio-visual modality fusion}
In this section,  we explain the details of three different modality fusion methods used in our AVSR models: feature concatenation, visual modality driven gated fusion, and audio-visual modality driven fusion.

\label{ssec:subhead}
\subsubsection{Feature concatenation based fusion}
\label{sssec:subhead}
The baseline AVSR system using feature concatenation is illustrated in Figure 2 (a). The acoustic features are concatenated with the visual features extracted by the LipNet, then the concatenated features are passed to the RecogNet:
\begin{align}
    p({\bf{y_{t}|x_{t}}})={\bf{RecogNet( [x_{t},LipNet(v_t)] )}},
\end{align}
where $\bf{y_t}$ is the frame-level alignment of the correspond acoustic frame $\bf{x_t}$, $\bf{v_t}$ is the mouth ROI of the target speaker.

\subsubsection{Visual modality driven gated fusion}
\label{sssec:subhead}
Since visual modality is invariant to the acoustic degradation, the gated architecture is purposefully designed to extract the target speaker from overlapped speech. Compared with concatenation, gating operation is more natural and direct for doing selection.  
Figure 2 (b) illustrates the structure of the visual modality driven gate. First, the acoustic features and the visual features are passed to ViusalNet and AudioNet networks respectively. Then, the outputs of the AudioNet are gated by the outputs of the VisualNet $m_{t}$ with an element-wise multiplication:
\begin{align}
    &{\bf{m_{t}}}={\bf{VisualNet(v_t),}}\\ \nonumber
    &{\bf{h_{t}}}={\bf{AudioNet(x_{t}),}}\otimes \sigma({\bf{m_{t}}}) \nonumber \\
    & p({\bf{y_{t}|x_{t}}})={\bf{RecogNet(h_t)}},\nonumber
\end{align}
where $\otimes$ denotes the Hadamard product, $\sigma(\cdot)$ is sigmoid function. The AudioNet and VisualNet have a similar architecture, each with 6 TDNN-F layers.

\subsubsection{Audio-visual modality driven gated fusion}
\label{sssec:subhead}
Although the visual modality is invariant to the acoustic degradation, it contains less information in terms of words or phones discrimination compared with the audio modality. This can easily be conformed by the significant performance gap between lipreading and ASR. Therefore, just relying on the visual feature for the gating process might not the most effective approach. In this case, as shown in Figure 2 (c), a so-called FusionNet is added to the visual modality driven gate structure. The outputs of the AudioNet and the VisualNet are concatenated and sent into the FusionNet before used in the gating step:
\begin{align}
    {\bf{m_{t}}}=\bf {FusionNet( [{\bf{VisualNet(v_t)}},{\bf{AudioNet(x_{t})}}] ),}
\end{align}
where the FusionNet is a TDNN network containing 3 tdnn-f layers. We expect that the FusionNet can leverage both the distorted audio and visual information to provide more effective information for the gating step.
%Then the outputs of the AudioNet will be gated by the output of the Fusion Net. The Fusion Net shares the same architecture with the Audio Net and the Visual Net. For the similar reason in section 2.2, the gated outputs are concatenated with visual features before fed into the RecogNet as an optional method in the two gating architectures. 

% It is worth to mention that though these two methods look similar to the so-called jointly training approaches, they are different from a jointly trained pipelined system for 1) using log filterbank features as input; 2) not producing any waveform or spectrum outputs; 3) not pretraining any separation parts using parallel overlapped and clean data.

\subsection{Integrated audio-visual system for overlapped speech}
In contrast to the pipelined system with explicit separation and recognition components shown in Figure 1 (a), the integrated system as shown in Figure 1 (b) tries to implicitly do both separation and recognition in a compact model architecture using a single recognition cost function. The architectures in Figure 3 (b) and (c) can be viewed as two integrated architectures for overlapped speech recognition. More specifically, in these two models, the front-end part including the gate structure can be regarded as an implicit speech separation component, and the RecogNet in the back-end can be seen as a recognition component. The entire model is trained to optimize the LF-MMI sequence training objective function. 
Moreover, as the dashed arrow in Figure 2 (b) and (c) shows, the gated hidden outputs are further concatenated with the visual features before passed into the RecogNet. The motivation is that the visual features is still complementary to the gated acoustic representations. 

% In addition, the dashed line in Figure 2 (b) and (c) denotes concatenating the gated hidden outputs with the visual features. Once you enhanced the acoustic, even the clean acoustic is also complementary to the visual, that motivates us to further concatenate the visual with enhanced acoustic.

% The structure in Figure 2 (b) and (c) can be regarded as integrated systems containing implicit separation and recognition components. Compared with the pipelined system having explicitly separation and recognition components trained by different cost functions, the integrated system is consistently optimized by the ASR cost function, i.e. LF-MMI. 

\section{Experiment \& Results}
\label{sec:typestyle}
Experiment setup is introduced in section 4.1, followed by the experiment results.

\subsection{Experiment setup}
\label{sssec:subhead}

\noindent{\bf Dataset:} In the experiment, we created two-speaker mixtures using utterances from Oxford-BBC Lip Reading Sentences 2 (LRS2) database \cite{chung2017lip}, which is one of the largest publicly available datasets for lip reading sentences in-the-wild.
The database consists of mainly news and talk shows from BBC programs. 
It is a challenging set since it contains thousands of speakers without speaker labels and large variation in head pose.
The dataset is already divided into training, validation (Train-Val) and test sets and also contains a pretraining (Pre-train) set with longer segments. 
% Details about the dataset can be found in Table 2. 
In this experiment Pre-train and Train-Val parts are combined as training set.

% \begin{table}[h]
% \centering
% \caption{ Statistics of the LRS2 dataset }
% \begin{tabular}{ccccc}
% \toprule
% Dataset &\#Utts  &\#Vocab &\#Hours\\ 
% \hline
% Pre-train &96k  &41k &195\\
% Train-Val &337k &18k &29\\
% Test   &1243  &1693 &0.5\\
% \bottomrule
% \end{tabular}
% \end{table}

\noindent{\bf Data simulation:} The overlapped speech utterances are generated by first sampling one reference audio-visual utterance from the training or test set and then mixing its audio with another interfering audio signals. To ensure the videos of each source are available in a mixture, longer sources are truncated to be aligned with the shortest one. We mix the training data with six level SNRs (15dB, 10dB, 5dB, 0dB, -5dB and clean), both the target and interference data are sampled from the training set. 
% Visual features are extracted from the sequence of image frames using a spatiotemporal residual network described in [23]. 
% The network contains a 3D convolution layer, followed by a common 18-layer 2D ResNet [24]. For every video frame it outputs a compact 512 dimensional feature vector. The frame rate of the video is 25 frames per second (fps), we upsample the visual features to 100 fps by duplicate each frame 4 times.

\noindent{\bf{Features:}} Log Mel-filterbank acoustic features with 40 bins are used, which are extracted with a 40ms window, 10ms hop-length at a sample rate of 16kHz. As for visual inputs, the mouth ROI of LRS2 is already centered, we further crop the center 112 by 112 pixel region of all video frames and up-sample them to 100 frame per seconds using linear interpolation.

\noindent{\bf{Model Architectures:}} Our speech recognition system is developed based on the Kaldi Toolkit. Since LRS2 doesn't have a dictionary, grapheme-state units are used in our experiment. 
A GMM-HMM model trained on LRS2 Train-val set is used to generate frame-level alignment. 
The alignment of the corresponding clean target speech is used for overlapped speech. 
All recognition systems are trained using the LF-MMI \cite{LFMMI} criterion using leaky HMM with the cross-entropy(CE) regularization.
The LipNet is pretrained on lipreading task similar to \cite{Afouras18b}. 
The details of the audio-visual separation model can be found in our previous work \cite{wu2019time}

\noindent{\bf{Language Model:}} The language model is a 4gram language model trained on the transcriptions of the LRS2 Pre-train set which contains more than 2 million words.

\subsection{Hybrid vs End-to-End of AVSR }
In this paper, we compare the performance of our hybrid LF-MMI TDNN system with previous end-to-end system: TM-CTC \cite{afouras2018deep}, TM-seq2seq \cite{afouras2018deep}, and hybrid CTC/Attention structure \cite{petridis2018audio} on LRS2 dataset. The structure of the hybrid system used in this experiment is shown in Figure 2 (a). Results in Table 1 illustrates that our LF-MMI TDNN system significantly outperforms CE trained TDNN system and the previous state-of-the-art end-to-end models in visual-only (lipreading) , audio-only and audio-visual speech recognition tasks. The performance of CE trained TDNN system stands in the middle of different end-to-end systems, the gain of the hybrid system is mainly come from the sequence training criterion. 
It is worth to mention that the visual only system is supervised trained using audio frame-level alignments, which implies the potential to improve the performance of lipreading models using audio information.
\begin{table}[h]
\footnotesize
\centering
\caption{ WERs of hybrid  and end-to-end systems on visual only, audio-only and audio-visual speech recognition tasks.}
\begin{tabular}{c|c|c|c}
\toprule
Models &V &A &A+V\\
\hline
TM-CTC \cite{afouras2018deep} &65.0  &15.3  &13.7  \\
TM-Seq2seq \cite{afouras2018deep} &49.8  &10.5 &9.4  \\
CTC/Attention \cite{petridis2018audio}  &63.5  &8.3  &7.0  \\
\hline
\hline
CE TDNN       & 55.02        & 10.17       & 8.95\\
LF-MMI TDNN   &{\bf{48.86}}  &{\bf{6.71}}  &{\bf{5.93}} \\
\bottomrule
\end{tabular}
\end{table}

\subsection{Resutls of pipelined AVSR }
\label{ssec:subhead}
Table 2 shows the pipelined systems' word error rates (WERs) on two-speaker overlapped speech recognition task under four SNR conditions: -5dB, 0dB, 5dB and 10dB. 
Both the audio-only and audio-visual separation model in the pipelined system are trained using two-speaker overlapped speech simulated from LRS2 dataset. The first four rows in Table 2 shows the results of the pipelined system using clean speech trained ASR and AVSR back-end.
A stronger pipelined system using ASR and AVSR trained on the mix of clean and enhanced (separated) data are also displayed in the last two rows. 
The separated data are obtained by feeding the overlapped training data into the audio-visual separation system. 
The results indicates that the use of visual information in front-end separation component,  back-end recognition component  or both of them can remarkably improve the performance of the pipelined system.

\begin{table}[h]
\footnotesize
\centering
\caption{ WERs on audio-only and audio-visual pipelined systems, 'mult' means using both clean and separated as multi-conditional training data}
\scalebox{0.9}{
\begin{tabular}{cc|ccc|ccccc}
\toprule
\multicolumn{2}{c|}{Separation} &\multicolumn{3}{c|}{Recognition} & \multicolumn{5}{c}{WER}     \\
A      &V   &Data   &A      &V       & 10dB  & 5dB   & 0dB   & -5dB  & AVE \\
\hline
% \multirow{4}{*}{Time}&\cmark& \xmark&  \multirow{4}{*}{clean}&\cmark&\xmark  &15.87	&18.57	&26.81	&40.24	&25.37   \\
% &\cmark& \xmark&  &\cmark&\cmark  &10.83	&11.42	&16.86	&24.31	&15.86 \\
% &\cmark& \cmark&  &\cmark&\xmark  &10.01	&11.92	&16.61	&25.26	&15.95  \\
% &\cmark& \cmark& & \cmark&\cmark  &\bf{7.86}	&\bf{8.71}	&\bf{11.14}	&16.30	&\bf{11.00} \\
% \hline
% \hline
\cmark& \xmark& \multirow{4}{*}{clean}& \cmark&\xmark  &21.26	&29.35	&40.79	&55.43	&36.71  \\
\cmark& \xmark&  &\cmark&\cmark  &14.38	&19.87	&27.54	&39.06	&25.21  \\
\cmark& \cmark&  &\cmark&\xmark  &12.74	&14.94	&21.52	&32.73	&20.48  \\
\cmark& \cmark&  &\cmark&\cmark  &9.70	&11.10	&15.36	&22.98	&14.79  \\
\hline
\hline
\cmark& \cmark&  \multirow{2}{*}{mult}&\cmark&\xmark  &10.43	&14.94	&15.88	&21.81	&15.06  \\
\cmark& \cmark&  &\cmark&\cmark  &8.15	&9.22	&11.44	&14.86	&\bf{10.92}  \\
\bottomrule
\end{tabular}}
\end{table}

% \begin{table}[h]
% \footnotesize
% \centering
% \caption{ WERs on time-domain and frequency-domain based pipelined systems, '$+$' means using both clean and separated training data}
% \scalebox{0.9}{
% \begin{tabular}{c|cc|cc|cccccc}
% \toprule
% \multirow{2}{*}{Domain}&\multicolumn{2}{c|}{Sep} &\multicolumn{2}{c|}{ASR} & \multicolumn{5}{c}{WER}     \\
% &A      &V      &A      &V       & 10dB  & 5dB   & 0dB   & -5dB  & AVE \\
% \hline
% \multirow{4}{*}{Time}&\cmark& \xmark&  \cmark&\xmark  &15.87	&18.57	&26.81	&40.24	&25.37   \\
% &\cmark& \xmark&  \cmark&\cmark  &10.83	&11.42	&16.86	&24.31	&15.86 \\
% &\cmark& \cmark&  \cmark&\xmark  &10.01	&11.92	&16.61	&25.26	&15.95  \\
% &\cmark& \cmark&  \cmark&\cmark  &\bf{7.86}	&\bf{8.71}	&\bf{11.14}	&16.30	&\bf{11.00} \\
% \hline
% \hline
% \multirow{4}{*}{Freq}&\cmark& \xmark&  \cmark&\xmark  &21.26	&29.35	&40.79	&55.43	&36.71  \\
% &\cmark& \xmark&  \cmark&\cmark  &14.38	&19.87	&27.54	&39.06	&25.21  \\
% &\cmark& \cmark&  \cmark&\xmark  &12.74	&14.94	&21.52	&32.73	&20.48  \\
% &\cmark& \cmark&  \cmark&\cmark  &9.70	&11.10	&15.36	&22.98	&14.79  \\
% \hline
% \hline
% \multirow{2}{*}{Freq}&\cmark& \cmark&  \cmark$^+$&\xmark  &10.43	&14.94	&15.88	&21.81	&15.06  \\
% &\cmark& \cmark&  \cmark$^+$&\cmark  &8.15	&9.22	&11.44	&14.86	&\bf{10.92}  \\
% \bottomrule
% \end{tabular}}
% \end{table}

\subsection{Results of integrated AVSR}
\label{ssec:subhead}
In Table 3, the first two rows are the results of the ASR and AVSR systems trained on clean speech. The rest of the systems are trained on the mixture of clean and two-speaker overlapped speech. 
The system performance can be significantly improved up to {\bf{29.98\%}} absolute WER reduction by adding visual modality.
We observed that the visual modality driven gated fusion (V$\otimes$A) and audio-visual modality driven gated fusion (AV$\otimes$A) methods significantly outperform the feature concatenation based method, which indicates efficacy of gating operation in overlap speech recognition. 
Compared with the pipelined systems results in Table 2, the best integrated system is slightly better than the best pipelined systems using multi-conditional trained AVSR. 
% We test the forward time of an 1 second utterance in (AV$\otimes$A) and best pipelined models on a single CPU, which are {\bf{0.4s}} and 0.47s respectively.
% It is worth to mention that time-domain system considers both the phase and amplitude information in the separation stage, while the frequency-domain and integrated system only use the amplitude information, which implies the potential to further improve the integrated system's performance by developing the using time-domain inputs or phase information.

\label{sssec:subhead}
\begin{table}[h]
\centering
\footnotesize
\caption{ WERs on integrated audio-visual overlapped speech recognition. 'concat' denotes feature concatenation fusion, "+concat" denotes concatenating the visual feature with the gated output. "mult*" denotes using clean and overlapped speech with different SNR level as training data }
\scalebox{0.9}{
\begin{tabular}{c|cc|c|ccccc}
\toprule
\multirow{2}{*}{Data} & \multicolumn{2}{c|}{Modality}&\multirow{2}{*}{Fusion}&&&WER\\
&A&V&	&10dB	&5dB	&0dB	&-5dB	&AVE     \\
\hline
\multirow{2}{*}{clean} & \cmark &\xmark&-	&19.01	&37.94	&66.78	&84.17	&51.98 \\
 &\cmark&\cmark&concat	&10.05	&20.81	&40.33	&63.87	&33.77 \\
\hline
\hline
\multirow{2}{*}{mult*}& \cmark&\xmark&-	&11.37	&21.67	&52.45	&75.68	&40.29  \\
&\cmark&\cmark&concat	&8.69	&11.29	&16.25	&24.58	&15.20   \\
\hline
\hline
\multirow{4}{*}{mult*} & \cmark&\cmark &V$\otimes$A &9.05	&10.47	&13.60	&17.17	&12.57 \\
 &\cmark&\cmark& \ +concat &8.23	&9.80	&12.52	&15.78	&11.49 \\
\cline{2-9}
 & \cmark&\cmark& AV$\otimes$A &7.87	&9.16	&11.42	&14.76	&10.80 \\
 & \cmark&\cmark& \ +concat &\bf{7.55}	&\bf{8.77}	&\bf{10.71}	&\bf{14.22}	&\bf{10.31}\\
\bottomrule
\end{tabular}
}
\end{table}

\section{Conclusion \& future work}
\label{sec:majhead}
This study first investigates the performance of LF-MMI trained hybrid AVSR and end-to-end AVSR systems on LRS2 dataset, a new state-of-the-art result is established by our LF-MMI TDNN model, then proposes two gated fusion methods purposely for overlapped speech recognition and compares the proposed methods with traditional pipelined systems. Experiments show: 1) the hybrid AVSR system outperforms end-to-end systems on LRS2 dataset; 2) the effectiveness of the gated fusion method; 3) the integrated system have comparable result with more complex pipelined system. In the future, this work will be extended to: 1) true cocktail party environment with noise, interference speech and reverberation; 2) a multi-channel system; 3) challenging situations, such as both the visual and audio information are degraded. Comparison between integrated system and jointly/multi-task training systems will also be investigated in the future.

% Below is an example of how to insert images. Delete the ``\vspace'' line,
% uncomment the preceding line ``\centerline...'' and replace ``imageX.ps''
% with a suitable PostScript file name.
% -------------------------------------------------------------------------

%\vfill
%\pagebreak

 \vfill\pagebreak

\bibliographystyle{IEEEbib}
\footnotesize{
\bibliography{strings,refs}
}

\end{document}